# GENERATIONAL FRAMESHIFTS IN TECHNOLOGY: COMPUTER SCIENCE AND NEUROSURGERY, THE VR USE CASE

Samuel R. Browd[1], Maya Sharma[2], Chetan Sharma[3]

[1]Department of Neurological Surgery, University of Washington, Seattle, Washington, USA, [2]Eastside Preparatory School, Kirkland, Washington, USA, [3]Chetan Sharma Consulting, Issaquah, Washington, USA

Contact Information: Dr. Samuel Browd, sbrowd@gmail.com. Maya Sharma, reachmayasharma@gmail.com. Chetan Sharma, chetan@ieee.org

***Abstract*** *– We are at a unique moment in history where there is a confluence of technologies which will synergistically come together to transform the practice of neurosurgery. These technological transformations will be all-encompassing, including improved tools and methods for intraoperative performance of neurosurgery, scalable solutions for asynchronous neurosurgical training and simulation, as well as broad aggregation of operative data allowing fundamental changes in quality assessment, billing, outcome measures, and dissemination of surgical best practices. The ability to perform surgery more safely and more efficiently while capturing the operative details and parsing each component of the operation will open an entirely new epoch advancing our field and all surgical specialties. The digitization of all components within the operating room will allow us to leverage the various fields within computer and computational science to obtain new insights that will improve care and delivery of the highest quality neurosurgery regardless of location. The democratization of neurosurgery is at hand and will be driven by our development, extraction, and adoption of these tools of the modern world. Virtual reality provides a good example of how consumer-facing technologies are finding a clear role in industry and medicine and serves as a notable example of the confluence of various computer science technologies creating a novel paradigm for scaling human ability and interactions. The authors describe the technology ecosystem that has come and highlight a myriad of computational and data sciences that will be necessary to enable the operating room of the near future.*

## 1. INTRODUCTION

The last twenty years have seen a complete shift from an analogue to a digital society. The shift to digital emcompasses all aspects of our lives, from how we shop, bank, learn, communicate, etc. There remain few vestiges of the analogue world in our lives, bar one location; the operating room.

A critical look shows technology rooted in ideas and methods that are decades, if not generations, dated. Microscopes fundamentally use ground and polished glass to magnify the scene. Navigation technologies are rooted in technologies that are decades old. Information gathering and sharing is archaic and remains largely analog. Along the same lines, learning and training have not changed appreciable since the inception of surgical apprecticeships. Even with the advent of digital books, the text is flat and video is largely 2-dimentional.

There are several reason that the digitization of the operating room has been slow to arrive. HIPAA regulations and our important and necessary drive to ensure patient privacy has limited movement of data outside of the surgeon's home instiution. Moving data from source to cloud and back introduces security risk and potential data breach but this said, parallel advancements in cryptography and security has transformed areas such as banking and other areas of the financial sector.

The moment has arisen when there is an intersection of technologies that are coming together that will fundamentally changes how surgery is taught, performed, and improved. Concepts, technologies and regulations forged in the consumer markets are being brought to the operating room in this unique moment of surgical history. We review some of the fundental technologies that will shape the modern surgical technology revolution and disucss VR as a multi-facet proxy to these enabling technologies.

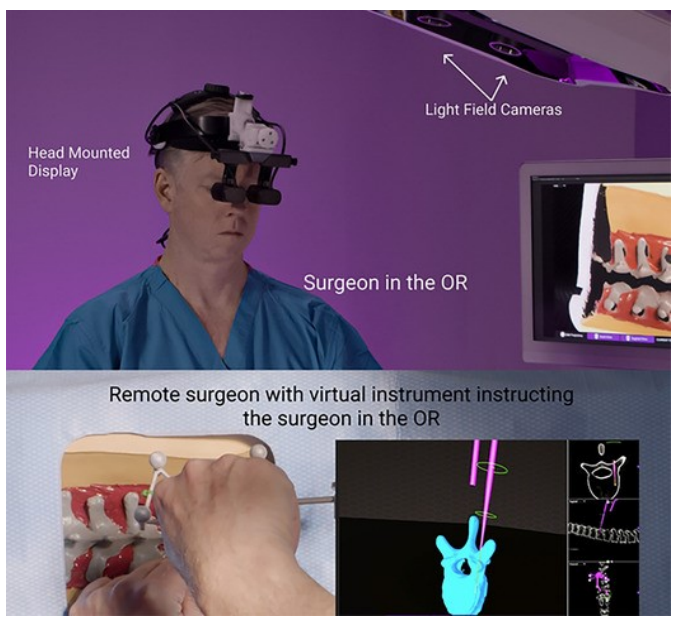

**Fig 1:** Remote VR surgery setup. (Top) In this case, one of the authors (SR) is shown with the VR HMD in the operating room working on the patient. (Bottom) View seen by the surgeon on the VR headset. The surgeon sees the actual imagery being transmitted through a communications network, and the images on the bottom right are being rendered in real time so the remote surgeon can interact with the live surgery using hand gestures like picking up the virtual screwdriver to indicate where exactly the incision needs to be made on the spine. The various elements of AI, Edge Computing, and 5G will help in making the system work end-to-end. (The authors were involved in demonstrating one of the earliest remote VR surgeries in September 2019 using the components and technologies discussed in this paper. Video of that demo is available at https://youtu.be/iwEeA0GzcZA.) © Propriovision.

As we look towards the next decade and what is coming, we see a significant number of new technologies becoming available as a toolset for transforming industries like healthcare and specifically surgery, and we appear on the verge of a full frameshift in enabling technologies. The development and integration of these new technologies are driven by multiple forces, including desired improvements in care, lower costs, higher quality, and equitable delivery of care and training. Virtual reality (VR) surgeries, as an example, will not be enabled by just one technology; instead, this emerging field will benefit from the growth of several adjacent and simultaneous technologies. VR will take advantage of artificial intelligence (AI), fifth-generation wireless network technology (5G), robotics, computational imaging, and edge computing, each of which is an emerging and rapidly growing area within computer and computational sciences that has blossomed within the last decade. These technologies have seen accelerated development and adoption in use cases outside of medicine. The intersection of these various technologies is creating a new technology s-curve in surgery (Figure 1) that will see the confluence of multiple technologies being brought into the operating room. VR is a fascinating example of how technology is evolving, and adaptations and new methods are being developed to bring these emerging technologies into the operating room.

## 2. ENABLING TECHNOLOGIES – STATE OF THE ART

Computer and computation devices no longer are singular platforms, driven by synclastic technologies. VR/MR/AR platforms of today incorporate myriad of functions from wireless functionality, to IT integration, to edge compute technologies. The digital OR will incorporate synergistic technologies such as 5G for data communications and access, AI and ML for data management and insights, robotics for automation, edge computing for data processing, and light-field imaging to digitize the visual field of view. Synergetic and combinatorial variations of such technologies will provide some of the biggest gains in health care. VR thus becomes, not a standalone technology, but requires leveraging new concomitant technologies to fully realize the value proposition inherent. For exampke, synthetic or semi-synthetic environments that enhance the surgeon's vision or situational awareness need to perform quickly, transfer relevant data efficiently and without error, and meet demanding physical requirements that will lead to embodiments of the technology that fit the ergonomic and workflow requirements in the operating room. Examples of enabling technologies that have impacted consumer and industries outside of surgery include 5G, AI/ML, edge compute, and visualization, experiential plaforms such as VR.

## 2.1   5G

As wireless connectivity accelerates (Figure 2), there is a new opportunity to seize upon the roll-out of 5G cellular-based technologies within healthcare, leveraging the revolutionary ability to move, share, and harvest data wirelessly from within the hospital and the operating room environment. The adaptation and integration of 5G will accelerate the use of wireless technologies in the OR including transfer of large data sets used in imaging as well as improve the ability to leverage cloud-based computational power. Head-mounted visual platforms that are integrated into the surgeons line-of-sight will increase in power and capacity while also, and importantly, undergoing continue improvements in ergonomic and weight. The wireless industry is going through a massive 5G network upgrade cycle [1][2].

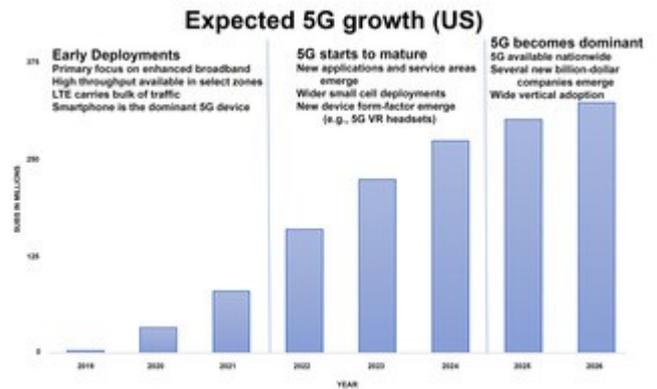

**Fig 2**. Expected growth of 5G in the U.S [3]

5G brings in a range of new features and capabilities that will be transformative for healthcare. In addition to gigabit speeds and sub 20-msec latencies, 5G offers features such as network slicing that can dedicate a slice of the network to an application, support distributed sensors nodes in a dense environment like the operating room, and provide enhanced security for data transmission. Further, the trends around having private networks to build a dedicated network for hospital campuses or operating rooms using cellular technologies are quite attractive because they reduce interference and improve reliability of wireless networks in hospital environments. The scale of 5G cellular handsets in the market will lead to network effects and lower the pricing for 5G chipsets such that they can be embedded in new forms of devices such as VR headsets, which will further accelerate the advent of new applications and services.

5G will impact surgery by allowing operating rooms to be configured without wires using 5G infrastructure that use high-band spectrum allowing for gigabit-per-second speed that is free of interference and provides extremely low latency. This will be critical for VR and head-mounted device data transmission in the operating room, rendering of high-resolution images, overlaying of real-time data within the synthetic surgical view, and facilitating telesurgical platforms that will require high-speed, low-latency data transfer to facilitate real-time interactions between surgeons positioned across the globe. 5G is attractive because it offers this high-bandwidth capability in higher bands coupled with significantly lower latency.

## 2.2 Artificial Intelligence (AI)

AI or automation of data analysis and intelligence is one of the most fundamental shifts that is occurring across all industry verticals. The role of AI in healthcare is going to be transformational because of the enormity and complexity of data that must be analyzed both in real-time as well as post-processing [4]. AI can help decipher data patterns much faster than even the most well-trained and experienced professionals and, as such, it will become a mainstay within surgery allowing for real-time decision-making assistive tools. The real superpower of AI comes from ingesting data and finding meaning in categorical, noncategorical, or contextual data that might take hours or months to find or might be unrecognized by the human eye. AI is the most basic layer in the software stack that induces a layered approach to intelligence-driven decision-making, whether it is detecting surgical anatomy, outlining surgical margins, or providing insights within the surgical environment in real-time, on-demand. AI can be trained to examine patterns within an individual or interrogate data of populations, surfacing meaningful trends, patterns, and learnings that would otherwise go undetected.

As an example, marrying AI to VR opens a host of interesting abilities. Understanding the context and phase of an operation allows for smart decision support that can be based on a growing repository of computation data related to best practices, surgical outcome metrics, and importantly intraoperative risk assessment and warnings such as approaching critical structures or passing anatomical planes that could lead to injury or poor outcome. Surgical decision support could offer suggestions related to aneurysm clip selection, pedicle screw size/length recommendations, final spinal construct alignment or assessment of electrocorticography and optimal surgical resection boundries; all happening in real time with true contextual knowledge. If we leverage the masters of neurosurgery to inform our practice, how could their knowledge and outcomes be transferred with contextual relevance to other surgeons and applied in a specific case at the right moment in the operation. AI delivered via VR synthetic vision could allow for the transfer of knowledge in the operating environment in real time, presented in a way that integrates the true anatomical view with the synthetic data. The digitization of the operating room including capture of the visual scene and performance of surgery allows us to capture, process and anaylize in a way that has not been possible. In aggregate, the pearls and wisdom of surgery can be capture, analyzed and shared making what was once tribal knowledge handed down one to the next; instead, widely and freely shared and made available and deployed in surgery at the right moment and in context.

## 2.3 Edge Computing

The Edge Internet is a computational overlay architecture that essentially upgrades the existing network infrastructure [5]. Edge computing allows for data processing at the device in use, without having to fully access the cloud to transmit, compute, and resend data back to the source device. Thus, edge computing allows for reduction in data latency times and creates the ability to quickly cycle data for specific use cases [6]. The importance in surgery to have rapid, precise data turn-around makes edge computing a logical and important advancement as these high-compute and communications technologies are positioned for patient care.

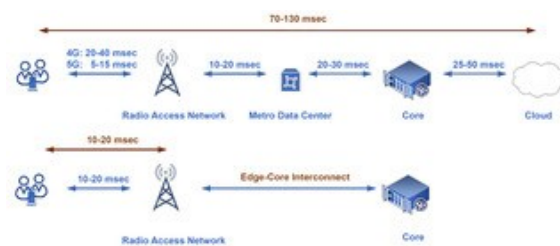

**Fig 3.** (Top) Legacy architecture; (bottom) Edge Internet architecture.

The data generation in hospitals is expected to reach over 3 Terabytes/day [7]. Obviously, only a small fraction of the data produced needs to be transmitted, but it needs to be processed, sometimes in real-time, and the current architecture is inadequate for handling the new demands and the new opportunities.

Figure 3 further expands on how such an architecture would work in a wireless network. In a VR-assisted neurosurgical environment, Edge computing capability is likely to be distributed between the VR headset, surgical equipment, 5G small cell, operator's radio access network, and data centers, depending on the workloads and latency requirements of a specific use case.

The requirements for computation and communications are enormous. Just doing some basic math, for VR, each of the human eyes can see up to 64 million pixels at any given moment, and with 120 frames per second (fps) requirement to generate a real-life view, we are looking at 15.5 billion of pixels/second. By storing each colored pixel in 36 bits and with 1:600 video compression available in H.265 HEVC encoding, we are looking at the consistency of 1 Gbps to guarantee acceptable user experience [8]. Additionally, 5-msec latency is required to ensure high-resolution and undistorted vision. These kinds of requirements are impossible to fulfill with legacy Internet and communications architectures.

As the discussion above illustrates, 5G, AI, and Edge computing come together to aid VR neurosurgery in a way that empower the neurosurgeon with data, insights, and visualization in real-time that will create better outcomes

## 3. VIRTUAL REALITY AND SURGERY

Within the last decade, the ability to synthetically create a virtual world environment has gone from science fiction to reality. Technological innovation in the virtual- and mediated-reality world has been occurring at breakneck speed, and the commoditization of VR devices has accelerated the use, adaption, accessibility, and cost of these devices. Although VR was originally considered for use primarily in the context of gaming and immersive player experiences, others quickly realized that the ability to synthetically create a novel world view could have implications throughout myriad contexts including business and industry. In the last five years, the medical community has also recognized the uses of VR modalities that look to address issues that can be uniquely dealt with using VR methodology. These use cases range from early examples in the study of anatomy to more recent use cases in the treatment of post-traumatic stress disorder, depression, anxiety and chronic pain. Surgeons quickly found utility in seeing three-dimensional representations of surgical anatomy in a format that allowed contextual information to be presented such that allowed the surgeon could move through and around the anatomy, manipulate the viewpoint and scale, and do some cursory renditions of operative procedures. The potential to early adopters has been obvious: VR stands to transform how information is presented to the surgeon before, during, and potentially after the operation.

The medical community has recently begun to recognize what has long been observed in the aviation industry, namely, that leveraging simulation for training and synthetic environmental immersion can offer great benefit. The ability to take learners and deploy them in a realistic training environment not only allows for the practice of routine skills, but also allows learners to experience challenging, anomalous, and dangerous situations in a risk-free environment. In medicine, this presents a safe situation that requires no direct patient interaction or risk. The scenarios can be repeatedly experienced and practiced, and the speed and ability to experience these situations is not limited by one's training experience, institution, or mentors. VR-based technologies will allow the practice sessions that have commonly happened in cadaver labs to be experienced at the time and place of the learner's choosing, fundamentally transforming how surgeons learn and experience new techniques, anatomical variants, and common intraoperative emergencies that until now are only experienced in the moment. As haptics and VR-augmented surgical tools are crafted for engaging in these training scenarios, the realism will only increase.

Beyond the training scenarios, VR will become integrated into operating procedures and can potentially be used by surgeons throughout the operation. Although these use cases are starting, they will initially be limited to value-added situations such as navigation. As VR and its enabling technologies that will become the digital operating room advance, VR will become the methodology of choice for many functions, including microscopy, navigation, data-directed surgeon augmentation, and directed decision-making. As the surgeon operates in VR (Figure 1), the relevant radiographic images will be superimposed in context and scale to the patient's anatomy; furthermore, the anatomic structures will be labeled, and suggestions will be presented to the surgeon regarding a variety of options as the case progresses. Examples might include which screw to select in fusion cases, which aneurysm clip is best suited for the specific aneurysm being clipped, and which wavelengths of light might be interpreted to assess and inform the surgeon as to the tumor margins.

VR will also enable telesurgery to occur in ways that, to date, have not been possible. VR headsets will be used to share the surgical experience in a fully immersive way with colleagues and trainees in the room, in the same city, and across the globe. The ability to bring in an expert who will be able to virtually participate in the case will transform the way expert surgeons are deployed and scaled for complex cases outside of their own catchment areas. Similarly, expert surgeons will be able to capture surgeries that can be annotated, archived, and shared broadly for training. The digitization of the operating room will enable new forms of education using VR including new textbooks that are immersive and interactive.

Although VR headsets are becoming more common and commoditized in the consumer space, associated technologies will need to be developed to bring the imaging and other enabling data in a usable format to the surgeon. The areas of 5G broadband communications, computer vision, data transfer and connectivity, AI and machine learning (ML), and edge computing will enable the digital operating room of the future and the ubiquitous integration of VR technologies there.

## 4. VR SURGERY ROADMAP

As has been established in the paper, VR technology will have a profound impact on how surgeries are done in the future. However, we must work through the limitations of the VR environment in the short-term and be cognizant of the influencing factors on the roadmap. For example, VR is highly sensitive to jitter, latency, and bandwidth and each deployment is going to be slightly different due to the radio-frequency characteristics of the operating room and the hospital, so, these performance KPIs must be measured for consistency and reliability. In scenarios where some surgeons might be joining the live surgery from remote, connectivity on the other end also becomes paramount. While the VR headsets are powerful computing device, the battery life is limited, and the headsets are bulky. It also takes some amount of getting used to VR in a surgical environment, so time needs to be carved out for training and testing. With the new sources of imaging, video, and VR data, data confidentiality elements of the policy and procedures need to be updated. New network nodes mean that such devices can be susceptible to hacks and as such need to be monitored and protected, data needs to be encrypted end-to-end, and data privacy laws need to be adhered to. Finally, given the introduction of a new technology, one has to update the legal liability frameworks as well.

## 5. OVERCOMING THE LIMITATIONS

Practitioners of the trade need to understand that future surgeries will be a joint IT and medicine undertaking. The medical expertise of surgeons will need to be ably assisted by VR surgery apparatus of edge computing, headsets, wireless networks, security, imaging, and more. Hiring of technical expertise at medical institutes will become essential to carry out VR surgeries. VR gives a completely new dimension to visualizing data and intelligence in real-time and it will take some time for surgeons to get used the new working environment. In fact, to reduce the learning curves, residents and medical students must be exposed to these new tools much early on so they develop proficiency in understanding the data, imaging, surgical references points and in trusting the tools. The technology space evolves faster than the procedures in medicine, so the teams need to stay up to speed on updates and upgrades across multiple technology areas such as the ones discussed in the paper. Finally, while initially the VR surgical apparatus might be expensive, the costs will come down as the market scales and the cost of equipment will easily be compensated by savings from simulation training and getting physicians better prepared, be safer and by reducing errors in the operating room.

## 6. CONCLUSION

Although originally developed in the gaming ecosystem, VR technology has found myriad uses in medicine across a variety of use cases, from post-traumatic stress disorder therapy to surgical planning. The potential VR applications will expand rapidly finding utility in many areas of surgical practice. VR allows for an immersive, shared experience that can happen in the same room or across continents. Because the data presented are synthesized, novel and unique applications can be brought directly into the operative environment. Surgeons can benefit from the overlay of data, imagery, shared intelligence, and decision support in real time as they prepare for and perform surgeries. Telesurgical collaborations will occur in real time, allowing colleagues to be visually transported into the remote operating room to participate in surgery with the same perspective and context as the physically present surgeon.

Technology cycles are happening quickly, and the integration of advanced computer science methodologies is creating new opportunities and disruptive trajectories within verticals such as surgery that will move the field forward into the next technological epic. VR, synthetic vision, and associated enabling technologies such as 5G, computer vision, VR, and AI will be transformational to the field of surgery, bringing sophisticated knowledge and decision support into the modern operating room. The evolution of the digital operating room, including VR, will advance the quality of care, reducing costs and facilitating training and education.

## ACKNOWLEDGEMENT

We thank Kristin Kraus, MSc, for editorial assistance.

**DISCLOSURES**

## AUTHORS

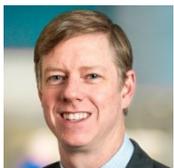

**Dr. Samuel R. Browd**, earned his M.D., Ph.D, from the University of Florida in 2000 and completed his Neurological Surgery residency at the University of Utah. He is a board certified pediatric neurosurgeon at Seattle Children's Hospital, Harborview and University of Washington Medical Center and a UW professor of Neurological Surgery and an adjunct professor of Biomedical engineering. He specializes in providing care to children with surgical disorders of the brain and spinal cord. He has published over 100 articles in peer-reviewed medical journals. Dr. Browd has also co-founded and served as chief medical officer of six companies in healthcare and holds multiple device patents in the fields of neurosurgery, computer science, and mechanical engineering. He currently serves as co-founder and chief medical officer of Propriovision, Inc.

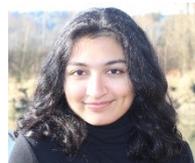

**Maya Sharma,** is a student at Eastside Preparatory School. Her research areas include application of new technology areas such as VR, 5G, IoT, Robotics, and AI to the medical field. She is the author of Paving – Conversations With Incredible Women Who Are Shaping Our World published by Olympia Publishers, London (2021). She is author of peer-reviewed journal papers in the International Youth Neuroscience Association Journal (IYNA), Health and Medicine related articles for The Sports Institute at University of Washingont School of Medicine, Op-Eds and articles related to medicine. She is a frequent speaker on STEM and leadership topics at conferences and on TV. Her essay on "Strategies and Solutions to solve Global Water Crises" won first place in statewide competition.

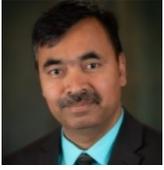 **Chetan Sharma,** received BS in Electrical Engineering from Indian Institute of Technology, Roorkee in 1991 and MS in EECE from Kansas State University in 1993. Since 2000, he has been CEO of a research and strategy consulting firm - Chetan Sharma Consulting. He is author or editor of 15 books related to wireless and has publsihed over 250 papers, articles, and reports on the wireless industry. His research interests are in applying new computing and communications technologies like 5G and Edge Computing to vertical industries.